\documentclass{ws-procs9x6}

\begin{document}

\title{ON THE AETHER-LIKE LORENTZ-BREAKING ACTION
FOR THE ELECTROMAGNETIC FIELD}

\author{J.R.\ NASCIMENTO and A.YU.\ PETROV$^*$}

\address{Departamento de F\'{\i}sica, Universidade Federal da Para\'{\i}ba\\
 Caixa Postal 5008, 58051-970, Jo\~ao Pessoa, Para\'{\i}ba, Brazil\\
$^*$E-mail: petrov@fisica.ufpb.br}

\author{M.\ GOMES and A.J.\ DA SILVA}
\address{Instituto de F\'\i sica, Universidade de S\~ao Paulo\\
Caixa Postal 66318, 05315-970, S\~ao Paulo, SP, Brazil}

\begin{abstract}
We study the  generation of  aether-like Lorentz-breaking actions in three, four and five dimensions via an appropriate Lorentz-breaking coupling of electromagnetic and spinor fields. Special attention is given to the four-dimensional case where we discuss the issue of ambiguities.
\end{abstract}

\bodymatter

\section{Introduction}

The breaking of Lorentz symmetry  has attracted great scientific attention. One of the more interesting direction in its study is the search for  possible Lorentz-violating extensions of field theory models. A review of possible forms of such extensions has been presented in Ref.\ \refcite{Kostel}. At the same time, we should note that the most known examples of the Lorentz-violating terms also break the CPT symmetry being constructed on the basis of a constant vector. Typical examples  are the Carroll-Field-Jackiw term \cite{CFJ0,CFJ} and the gravitational Chern-Simons term \cite{JaPi}.

Thg CPT-odd terms are not the only possible Lorentz-breaking terms; there are also  CPT-even Lorentz-breaking terms. In this context, some examples were discussed \cite{KostColl2} in the analysis of the Lorentz-violating Standard-Model Extension. The most important example of such terms is the aether term originally arising in studies of extra dimensions  \cite{Carroll}.  Different aspects of the aether terms for the scalar, gauge and spinor fields have been carried out in Ref.\ \refcite{ouraether}. In this communication, we discuss in detail the aether term for the electromagnetic field, particularly the problem of ambiguities.

\section{Aether term in electrodynamics}
The starting point of our study is the following action for the spinor field coupled to the electromagnetic field in a nonminimal manner:
\begin{eqnarray}
\label{edlb1}
S=\int d^Dx  \Big[\bar{\psi}(i\gamma\cdot\partial-m-
\tilde{\epsilon}^{abc}b_{a}F_{bc})\psi-\frac{1}{4}F_{ab}F^{ab}\Big],
\end{eqnarray}
where $\tilde{\epsilon}^{abc}$ is a matrix-valued object,
antisymmetric with respect to its Lorentz indices and defined as 
$\tilde{\epsilon}^{abc}=\epsilon^{abc}$ in $D=3$; $\tilde{\epsilon}^{abc}\equiv
\epsilon^{abcd}
\gamma_{d}$ in $D=4$, and
$\tilde{\epsilon}^{abc}=\epsilon^{abcde}\sigma_{de}$ in $D=5$.
Here $b_{a}$ is a vector implementing the Lorentz symmetry breaking. 
We note that in  three- and five-dimensional spacetimes the
one-loop contribution is finite within the framework of the dimensional 
regularization.

\begin{figure}
\begin{center}
\psfig{file=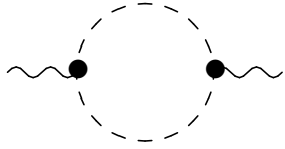,width=1.5in}
\end{center}
\caption{Two-point function of the gauge field.}
\label{aba:fig1}
\end{figure}

The lower order contribution to the two-point vertex function of the gauge field is shown in Fig.\ \ref{aba:fig1}.
Here the wavy line is for the external $F_{ab}$ field, the dark circle is for the Lorentz-breaking vector $b_a$ insertion, and the dashed line denotes the standard propagator of the spinor field $S(k)\equiv$ $<\bar{\psi}(-k)\psi(k)>={i(\gamma\cdot k+m)}/{(k^2-m^2)}$.

The contribution of this diagram gives
\begin{eqnarray}
\label{t1e}
S_2=\frac{g^2}{2}b_aF_{bc}b_{a'}
F_{b'c'}\int\frac{d^Dk}{(2\pi)^D}{\rm
  tr}\bigl[\,\tilde{\epsilon}^{abc}S(k)
\tilde{\epsilon}^{a'b'c'}S(k)].
\end{eqnarray}
Here we took into account that the external stress tensors can be treated as constants since the aether-like term does not depend on their derivatives.

In three and five dimensions this contribution is explicitly finite within the dimensional regularization scheme. Evaluating the integrals and matrix traces, one can arrive at the following expressions: in three-dimensional spacetime,
\begin{eqnarray}
S_2^{(D=3)}=\frac{2g^2}{\pi}|m|(b^aF_{ab})^2,
\end{eqnarray}
and in five-dimensional spacetime,
\begin{eqnarray}
S_2^{(D=5)}=-\frac{4g^2}{3\pi^2}|m|^3(b^aF_{ab})^2.
\end{eqnarray}
Thus, these contributions are proportional to $(b^aF_{ab})^2$, that is, to the aether-like term considered in Ref.\ \refcite{Carroll}, which is  a particular form of the CPT-even Lorentz-breaking term $k^{abcd}F_{ab}F_{cd}$ introduced in
Ref.\ \refcite{Kostel}.

\section{Four-dimensional case}

Now, let us concentrate on the four-dimensional case which requires  a more detailed discussion. In this case, the explicit form of the one-loop contribution (\ref{t1e}) looks like
\begin{eqnarray}
S_2(p)&=&-\frac{g^2}{2}\epsilon^{abcd}\epsilon^{a'b'c'd'}b_aF_{bc}(p)b_{a'}
F_{b'c'}I_{dd'},
\end{eqnarray}
where
\begin{eqnarray}
I_{dd'}=\int\frac{d^4k}{(2\pi)^4}\frac{1}{[k^2-m^2]^2}
{\rm tr}
\big[m^2\gamma_d\gamma_{d'}+k^mk^n\gamma_m\gamma_d\gamma_n
\gamma_{d'}\big].
\end{eqnarray}
This expression can be evaluated in several ways.

{\bf In the first way}, we carry out the symmetrization in four dimensions and only afterwards promote the integral to $d$ dimensions. It implies in the  replacement $k^mk^n\to\frac{1}{4}\eta^{mn}k^2$ with $\eta^{mn}\eta_{mn}=4$. After promoting the integral to $d=4+\epsilon$ dimensions and performing the integration we arrive at
\begin{eqnarray}
I_{dd'}=2i\eta_{dd'}\frac{\mu^{-\epsilon}}{(4\pi)^{2+\epsilon/2}}(m^2)^{1+\epsilon/2}[\Gamma(-\frac{\epsilon}{2})+\Gamma(-1-\frac{\epsilon}{2})]=-2i\eta_{dd'}\frac{m^2}{16\pi^2}. \quad
\end{eqnarray}
This expression is finite. One can show that this result is also reproduced using  the cutoff regularization and proper-time method.

{\bf In the second way}, we first promote the integral to $d$ dimensions and replace $k^mk^n\to\frac{1}{d}\eta^{mn}k^2$ with $\eta^{mn}\eta_{mn}=d$. In this case, after a Wick rotation, we have
\begin{eqnarray}
I_{dd'}=2i\eta_{dd'}\mu^{-\epsilon}\int\frac{d^dk_E}{(2\pi)^d}\frac{1}{[k^2_E+m^2]^2}
\big[\frac{2(d-2)}{d}k^2_E+2m^2\big].
\end{eqnarray}
By integrating, we arrive at
\begin{eqnarray}
I_{dd'}=2i\eta_{dd'}\frac{\mu^{-\epsilon}}{(4\pi)^{2+\epsilon/2}}(m^2)^{1+\epsilon/2}[\frac{2(2+\epsilon)}{4+\epsilon}\Gamma(-1-\frac{\epsilon}{2})+\frac{4}{4+\epsilon}\Gamma(-\frac{\epsilon}{2})].
\end{eqnarray}
By taking the limit  $\epsilon \to 0$, we obtain $I_{dd'}=0$. Therefore we conclude that the two-point function of the gauge field in the four-dimensional case depends on the regularization scheme. A similar situation has been observed for the CFJ term \cite{ourqed}. 

In the first case, when $I_{dd'}\neq 0$, we  carry out all contractions, so that  the two-point function for the electromagnetic term takes the form
\begin{eqnarray}
\label{s2d4}
S_2^{(D=4)}(p)&=g^2\frac{m^2}{4\pi^2}(b^aF_{ab})^2,
\end{eqnarray}
which reproduces the structure of the aether-like term  \cite{Carroll}. 

\section{Summary}

We have obtained the explicit  form of the aether-like term for the electromagnetic field. A peculiarity of our result consists in the fact that it is finite within the dimensional regularization approach in all spacetime dimensions from three to five. The finiteness of the aether-like contribution in the four-dimensional case is especially interesting since this contribution is superficially quadratically divergent. At the same time, while at least three methods of calculating this correction yield the same result, there is nevertheless one method which gives a different result. 
So, it would be interesting to know if an anomaly exists which could be responsible for the ambiguity of the four-dimensional aether term. This question is still open.

The CPT-even aether-like terms exist also for other fields. Their explicit expressions were found in all spacetime dimensions from three to five for the scalar and spinor fields. Therefore a relevant issue here is whether an aether-like term can arise as a quantum  correction in the gravity theory. This question will be considered  in the continuation of this study.

\section*{Acknowledgments}
The work by A.Yu.\ P.\ is supported by the CNPq project 303461-2009/8.

\end{document}